\documentstyle[amssymb]{aipproc}
\begin{document}
\pagestyle{plain}
\righthead{\sc Two Aspects of Deformation Programme}
\lefthead{\sc Mosh\'e Flato}
\setlength{\textwidth}{154mm}
\setlength{\textheight}{216mm}
\title{Two disjoint aspects of the deformation programme:
quantizing Nambu mechanics; singleton physics\footnote
{{\it This very short survey is dedicated to the memory of our good friend
Ryszard R\c aczka}, to whom the April 1998 meeting in \L \'od\'z ``Particles,
Fields and Gravitation", was dedicated. Except for this footnote 
the e-print is the August 1998 version to be published by AIP Press in
the Proceedings entitled {\it Particles, Fields and Gravitation}, edited by
Jakub Rembieli\'nski.}}
\author{Mosh\'e Flato}
\address{Laboratoire Gevrey de Math\'ematique Physique, CNRS \ ESA 5029\\
D\'epartement de Math\'ematiques, Universit\'e de Bourgogne\\
BP 400, F-21011 Dijon Cedex France \\ {\rm e-mail:}
{\tt flato@u-bourgogne.fr}}
\date{}
\maketitle
\vspace{-14mm}

\begin{abstract}
We present briefly the deformation philosophy and indicate, with references,
how it was applied to the quantization of Nambu mechanics and to particle
physics in anti De Sitter space.
\end{abstract}
\vspace{-7mm}

Deformation theory of algebraic structures has proved itself extremely
efficient and very much so in the last two decades. In this short contribution
(in fact, a long abstract with references) we shall indicate two recent
examples of new applications of the deformation philosophy. But before
doing so let us present the essence of our deformation philosophy
\cite{Fl82}.

Physical theories have their domain of applicability mainly depending on
the velocities and distances concerned. But the passage from one domain
(of velocities and distances) to another one does not appear in an 
uncontrolled way. Rather, a new fundamental constant enters the modified
formalism and the attached structures (symmetries, observables, states,
etc.) {\it deform} the initial structure; namely, we have a new structure
which in the limit when the new parameter goes to zero coincides
with the old formalism. In other words, to {\it detect} new formalisms we
have to study deformations of the algebraic structures attached to a given
formalism. 

The only question is in which category we perform this research of
deformations. Usually physics is rather conservative and if we start e.g.
with the category of associative or Lie algebras, we tend to deform in
this category. However recently examples were given in the literature
of generalizations of this principle. For instance quantum groups \cite{Dr86}
are in fact deformations of Hopf algebras. Other examples include more general
deformations like in the quantization of Nambu mechanics or in non Abelian
deformations \cite{Pi97}. Here we shall point out two recent developments
based on our deformation philosophy.

The first example has to do with the generalization of classical mechanics
suggested by Nambu in 1973 \cite{Nb73} and the recent attempts to quantize
it. The question there, which has to do with generalized deformations, is 
of general interest. Apparently Nambu mechanics presents a new alternative
to classical mechanics (see however \cite{BF75}). Does it possess a 
quantum version attached to it and what price do we have to pay to have 
such a version? The question is partially answered in \cite{DFST97,DF97} and
a detailed presentation can be found in \cite{FDS97}.

The second example has to do with recent developments in field theories
based on supergravity, conformal field theories, compactification of 
higher dimensional field theories, string theory, M-theory, $p$-branes,
etc. for which people rediscovered the efficiency and advantages of
anti De Sitter theories (which are stable deformations of Poincar\'e field
theories in the category of Lie groups; see however \cite{FHT93} for
quantum groups, at roots of unity in that case). There are many reasons
for the advantages of anti De Sitter (often abbreviated as AdS) theories
among which we can mention that AdS field theory admits an invariant
natural infrared regularization of the theories in question and that the
kinematical spectra (angular momentum and energy) are naturally discrete.
But in addition AdS theories have a great bonus: the existence of 
{\it singleton} representations discovered by Dirac \cite{Dir} for 
${\rm{SO}}(3,2)$, corresponding to a ``square root" of AdS massless
representations. We discovered that fact around 20 years ago \cite{FF78}
and developed rather extensively its physical consequences in the following
years \cite{FHT93,AF78,AFFS81,BFFH85,BFFS82,BFH83,FF88,FFxx,FF98,FFG86,%
FFS88,FaF80,Fr88,FH87,HFF92}.

Singleton theories are topological in the sense that the corresponding
singleton field theories live naturally on the boundary at infinity of the
De Sitter bulk (boundary which has one dimension less than the bulk).
They are new types of gauge theories which in addition permit to consider
massless particles, e.g. the photon, as {\it dynamically} composite
AdS particles \cite{FF88,FF98,FFS88}. Some of the beautiful properties of
singleton theories can be extended to higher dimensions, and this is the
main point of the recent huge interest in these AdS theories, which touched
a large variety of aspects of AdS physics. More explicitly, in several of
the recent articles among which we can mention 
\cite{Ma97,Wi98,FFr98,FFZ98,FZ98,FKPZ,FMMR,SS98}, 
the new picture permits to study duality between CFT on the boundary at
infinity and the corresponding AdS theory in the bulk. That duality, which
has also interesting dynamical aspects in it, utilizes among other things
the great notational simplifications permitted by singleton physics.

\vspace{-3mm}
{\footnotesize
\noindent {\textbf{PACS (1998)}}: 11.15.-q,11.25.Hf,11.30.Pb, 
12.20.-m,12.60.Rc, 14.80.-j, 04.62.+v,04.65.+e

}
\end{document}